\documentclass[aip,amsmath,amssymb,reprint,citeautoscript]{revtex4-1}

\usepackage{graphicx,dcolumn,bm,mathptmx,etoolbox,xcolor}
\usepackage{soul}
\usepackage[utf8]{inputenc}
\usepackage[T1]{fontenc}

\DeclareUnicodeCharacter{2215}{}

\makeatletter
\def\@email#1#2{%
 \endgroup
 \patchcmd{\titleblock@produce}
  {\frontmatter@RRAPformat}
  {\frontmatter@RRAPformat{\produce@RRAP{*#1\href{mailto:#2}{#2}}}\frontmatter@RRAPformat}
  {}{}
}%
\makeatother

\def\bte{\textit{ai}BTE}
\def\cms{cm$^2$/Vs}
\def\ganz{GaN@ZnGeN$_2$}
\def\ganm{GaN@MgSiN$_2$}

\def\oden{Oden Institute for Computational Engineering and Sciences, The University of Texas at
Austin, Austin, Texas 78712, USA}
\def\physics{Department of Physics, The University of Texas at Austin, Austin, Texas 78712, USA}
\def\ucsb{Materials Department, University of California, Santa Barbara, California 93106-5050, USA}
\def\epfl{Theory and Simulation of Materials (THEOS), \'Ecole Polytechnique F\'ed\'rale de Lausanne, CH-1015 Lausanne, Switzerland}
\def\uclouvain{Institute of Condensed Matter and Nanosciences, Universit\'e catholique de Louvain, Chemin des \'Etoiles 8, B-1348 Louvain-la-Neuve, Belgium}

\begin{document}

\title
[Anisotropic-strain-enhanced hole mobility in GaN by lattice matching to ZnGeN$_2$ and MgSiN$_2$]
{Anisotropic-strain-enhanced hole mobility in GaN by lattice matching to ZnGeN$_2$ and MgSiN$_2$}

\author{Joshua Leveillee}\affiliation{\oden}\affiliation{\physics}
\author{Samuel Ponc\'e}\affiliation{\epfl}\affiliation{\uclouvain}
\author{Nicholas L. Adamski}\affiliation{\ucsb}
\author{Chris G. Van de Walle}\affiliation{\ucsb}
\author{Feliciano Giustino}\affiliation{\oden}\affiliation{\physics}\email{fgiustino@oden.utexas.edu}

\date{\today}

\begin{abstract}
The key obstacle toward realizing integrated gallium nitride (GaN) electronics is its low hole mobility.
Here, we explore the possibility
of improving the hole mobility of GaN via epitaxial matching to II-IV nitride materials that have
recently become available, namely ZnGeN$_2$ and MgSiN$_2$. We perform state-of-the-art calculations of
the hole mobility of GaN using the \textit{ab initio} Boltzmann transport equation.
We show that effective uniaxial compressive strain of GaN along the $[1\bar{1}00]$ by lattice matching to ZnGeN$_2$ and MgSiN$_2$ results in the inversion of the heavy hole band and split-off hole band, thereby lowering the effective hole mass in the compression direction.
We find that lattice matching to ZnGeN$_2$ and MgSiN$_2$ induces an increase of the room-temperature hole mobility by 50\% and 260\% as compared to unstrained GaN, respectively.
Examining the trends as a function of strain, we find that the variation in mobility is highly nonlinear;
lattice matching to a hypothetical solid solution of Zn$_{0.75}$Ge$_{0.75}$Mg$_{0.25}$Si$_{0.25}$N$_2$ would already increase the hole mobility by 160\%.
\end{abstract}

\maketitle

Wurtzite gallium nitride has become a fundamental semiconductor component in a variety of electronic and optical devices, including radio-frequency applications\cite{Runton:2013,Zhang:2021}, power electronics\cite{Amano:2018}, and light emitting diodes (LEDs)\cite{Nakamura:1993,Nakamura:1995,Li:2016}.
Many remarkable properties of GaN can be traced back to its electronic structure.
As a wide-gap semiconductor with a band gap of 3.4 eV at room temperature\cite{Barker:1973,Madelung:1991}, GaN can support higher voltages without experiencing field-induced breakdown when compared with Si.
GaN can be alloyed with other nitride semiconductors such as AlN\cite{Wu:2004,Taniyasu:2006} and InN\cite{Arif:2007,Neufeld:2008} to tune the band gap between 0.6 and 6.2~eV, allowing for the realization of an array of materials for optoelectronic applications.

Electrons in GaN exhibit high room-temperature mobility, reaching up to 1000~cm$^2$/Vs\cite{Nakamura:1992,Gotz:1996,Gotz:1998,Kyle:2014,Ponce:2019,Ponce:2019_2,Jhalani:2020} at low defect concentration.
Conversely, hole carriers in GaN exhibit comparatively low mobility, usually below 31~cm$^{2}$/Vs\cite{Rubin:1994,Kozodoy:1998,Kozodoy:2000,Arakawa:2016,Horita:2017,McLaurin:2007}.
The imbalance between electron and hole mobility hinders the application of GaN to integrated electronics based on complementary field-effect devices, and severely limits the uses of $p$-channel GaN in power electronics and radio-frequency switching\cite{Amano:2018,Bader:2020}.
In order to meet these technological demands, materials engineering approaches aimed at increasing the hole mobility of GaN are needed.

\begin{figure*}
\includegraphics[width=1.8\columnwidth]{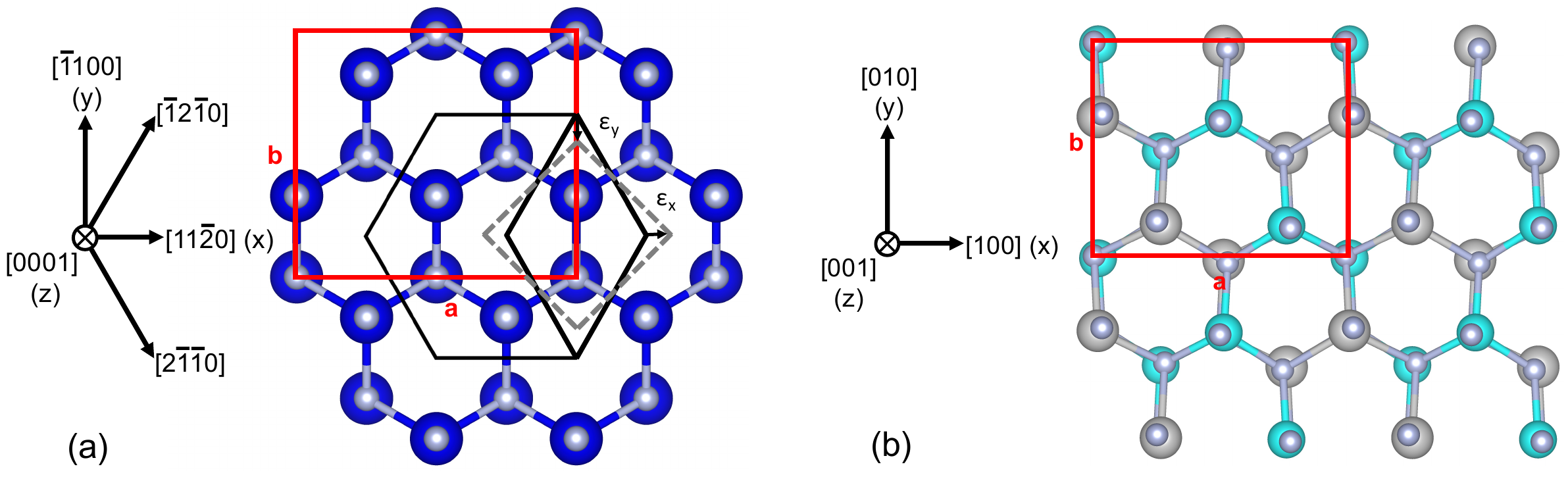}
\caption{\label{fig:compare_vecs}
(a) Crystal structure of GaN, with Ga shown in dark blue and N in silver. The primitive unit cell of GaN
is shown as the thick black diamond, and the orthorhombic unit cell is shown in red. The lattice parameters
$a$ and $b$ refer to the orthorhombic cell. The deformed gray diamond shows the strain directions (not in
scale). We also show how the Cartesian reference frame is aligned with respect to the standard crystallographic
directions of the hexagonal lattice. (b) Crystal structure of ZnGeN$_2$ and MgSiN$_2$, with Zn or Mg shown in
light blue, and Ge or Si in grey. The crystallographic directions of the orthorhombic lattice are aligned with
the Cartesian axis.}
\end{figure*}

A common approach to improving the carrier mobility of semiconductors is via strain engineering.
In the case of GaN, several experimental and theoretical/computational reports explored this possibility.
Using $\mathbf k \cdot \mathbf p$ perturbation theory, Suzuki \emph{et al.} found that compressive uniaxial strain along the in-plane $[\bar{1}100]$ axis (see Fig.~\ref{fig:compare_vecs}) induces the inversion of the heavy hole and split-off hole bands\cite{Suzuki:1996}.
The resulting valence-band maximum has a small effective mass along the $k_{y}$ direction, leading to increased hole mobility.
Using an effective-mass-theory approach, Yeo \emph{et al.} found that for growth in the (10$\bar{1}$0) crystal orientation, strain induced in a GaN quantum well sandwiched between AlGaN layers would lead to a lighter effective mass along the compression direction\cite{Yeo:1998}.
In recent experiments, Gupta \emph{et al.} realized uniaxial compressive strain of GaN in the basal plane of the wurtzite structure (perpendicular to the $[0001]$ axis, see Fig.~\ref{fig:compare_vecs}) by using a fin geometry, while allowing for strain release in the other directions\cite{Gupta:2019}.
This setup achieved a 25\%-50\% reduction in sheet resistance of $p$-type GaN along the compressive uniaxial strain direction.
Using the \textit{ab initio} Boltzmann transport equation (\bte), Ponc\'{e} \emph{et al.} showed that imposing tensile biaxial strain in the (0001) plane, resulting in 2\% compression along the $c$ axis, raises the split-off band above the heavy- and light-hole bands, thereby increasing the hole mobility along the same direction by 250\%\cite{Ponce:2019,Ponce:2019_2}.

In this work, we explore an alternative strategy for improving the hole mobility of GaN via strain engineering.
A possible route to inducing in-plane uniaxial strain perpendicular to the GaN crystal $c$ ([0001]) axis is via epitaxial growth of GaN on II-IV nitride materials with mismatched lattice parameters.
Two candidate materials for this purpose are ZnGeN$_2$ and MgSiN$_2$, which crystallize in an orthorhombic lattice with space group Pna2$_1$.
To discuss the relation between the lattice parameters of ZnGeN$_2$, MgSiN$_2$, and GaN, we use
a common reference frame for the hexagonal and the orthorhombic lattices, as shown in Fig.~\ref{fig:compare_vecs}.
The correspondence between hexagonal and orthorhombic directions is provided in Table~S1.

For ZnGeN$_2$, a range of values has been reported for the room-temperature (300 K) lattice parameters along the $[100]$ and $[010]$ directions [see Fig.~\ref{fig:compare_vecs}(b)]: $a=$6.38-6.45~\AA\ and $b=$5.46-5.52~\AA~\cite{Blanton:2017,Tellekamp:2020}.
We consider the values $a=$6.425~\AA\ and $b=$5.478~\AA\ to be representative as averages over the experimentally measured lattice parameters and will proceed to use these to determine strain values.
For MgSiN$_2$, accurate x-ray and neutron diffraction measurements at room temperature indicate lattice constants $a=6.473$~\AA\ and $b=5.272$~\AA~\cite{Bruls:2000}.


The lattice parameters of GaN in the orthorhombic cell outlined in Fig.~\ref{fig:compare_vecs}(a) are $a$ = 6.364~\AA\ and $b$ = 5.511~\AA~\cite{Gian:1996,Tellekamp:2020}, therefore the epitaxial matching of GaN to ZnGeN$_2$ and MgSiN$_2$ will induce a tensile strain $\epsilon_x$ along the $a$ ($x$) direction, and a compressive strain $\epsilon_y$ along the $b$ ($y$) direction.
For calculations of strained GaN, we impose the strain using the experimental room-temperature lattice parameters of GaN and the II-IV nitrides (see details in the supplementary material).
The resulting strain values are listed in Table~\ref{tab:strains}.
Growth of ZnGeN$_2$ on GaN has recently been demonstrated via molecular beam epitaxy~\cite{Tellekamp:2020}, paving the way for the realization of devices based on strained GaN.
Growth of GaN on MgSiN$_2$ or MgSiN$_2$ on GaN has not yet been explored. The effect of in-plane uniaxial strain on the mobility of GaN has been investigated theoretically using  a variety of effective-mass models\cite{Rode:1995,Look:1997,Look:1997_2,Abdel-Motaleb:2005}, but these models predate recent developments in predictive \textit{ab initio} calculations of carrier mobilities\cite{Ponce:2018,Ponce:2019,Jhalani:2020}, hence a study using state-of-the-art computational techniques is warranted.

We perform density functional theory (DFT) calculatons, density functional perturbation theory (DFPT) calculations, and
Wannier-Fourier interpolation of electronic structure, phonon dispersions, and
electron-phonon coupling matrix elements using the Quantum ESPRESSO package\cite{Giannozzi:2017},
the Wannier90 code\cite{Mostofi:2014}, and the EPW code\cite{Giustino:2007,Noffsinger:2010,Verdi:2015,
Ponce:2016}. We use the local density approximation (LDA)\cite{Ceperley:1980,Perdew:1981} and
optimized norm-conserving Vanderbilt (ONCV) pseudopotentials\cite{Hamann:2013,Schlipf:2015}.
In order to obtain accurate band structures and effective masses, we employ the GW method as implemented
in the Yambo code\cite{Marini:2009,Sangalli:2019,Godby:1989}, including spin-orbit coupling in the calculations.
The calculation of carrier drift mobility is performed using the \textit{ab initio} Boltzmann transport equation\cite{Li:2015,Madsen:2018} (\bte) and ultra-dense grids of wavevectors in the Brillouin
zone\cite{Ponce:2018,Ponce:2020,Ponce:2021,Brunin:2020}, as implemented in the EPW code.
A detailed description of the computational setup is provided in the supplementary material.

We perform structural optimization of the unit cell parameters and internal coordinates of unstrained GaN.
For strained GaN, we determine the strain configuration in the $xy$ plane using the experimental lattice parameters of GaN, ZnGeN$_2$, and MgSiN$_2$, and we optimize the $c$ parameter as well as the internal coordinates, see Fig.~\ref{fig:compare_vecs}.
Detailed information about the structural parameters and strain levels considered is reported in Table~S1 and Fig.~S1.
In the following we discuss three systems: unstrained GaN, and GaN strained to match ZnGeN$_2$ (GaN@ZnGeN$_2$) or MgSiN$_2$ (GaN@MgSiN$_2$) (Table~\ref{tab:strains}).

\begin{table}
\caption{\label{tab:strains} Strains along the $x$ and $y$ directions resulting from lattice matching GaN to ZnGeN$_2$ and MgSiN$_2$. }
\begin{ruledtabular}
\begin{tabular}{c c c}
     & $\varepsilon_x$ & $\varepsilon_y$ \\
     \hline
  \ganz \cite{Blanton:2017,Tellekamp:2020}   & 0.96\% & -0.60\% \\
  \ganm \cite{Bruls:2000}   & 1.72\% & -4.33\% \\
\end{tabular}
\end{ruledtabular}
\end{table}

\begin{figure*}
\includegraphics[width=2.0\columnwidth]{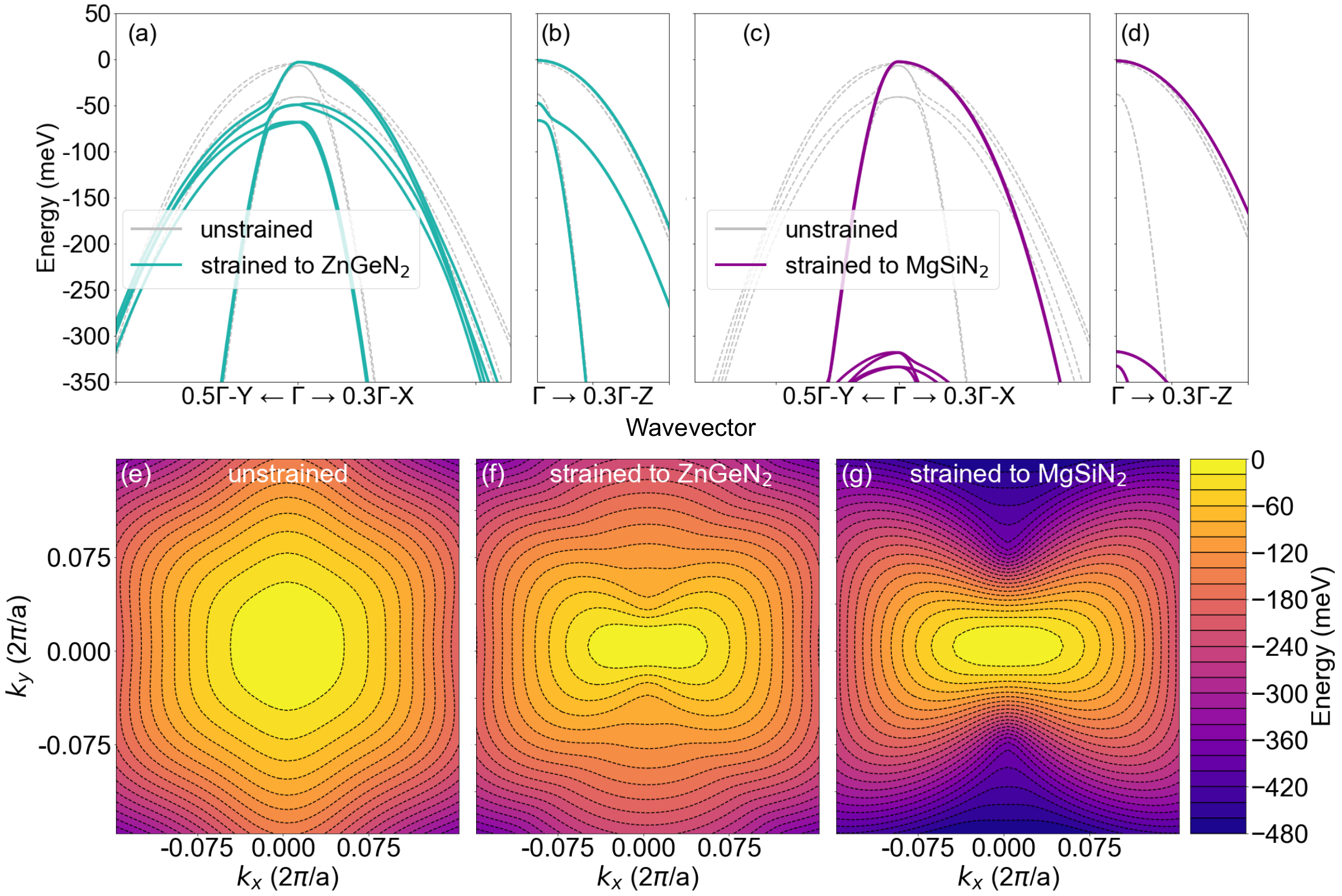}
\caption{\label{fig:bands}
(a) and (b) Valence-band structure of GaN (grey dashed lines) and \ganz\ (green solid lines) near the
band extremum. (c) and (d) Valence-band structure of GaN (grey dashed lines) and \ganm\ (purple solid lines)
near the band extremum. (e) Energy of the highest valence band of GaN vs.\ $k_x$ and $k_y$, in a
two-dimensional plot with $k_z=0$. (f) Same setup as in (e), but for \ganz. (g) Same setup as in (e),
but for \ganm.}
\end{figure*}

Figures~\ref{fig:bands}(a)-(d) show the valence-band structure of unstrained and strained GaN
near the band extremum. In panels (a) and (b) we compare unstrained GaN with GaN@ZnGeN$_2$,
and in panels (c) and (d) we compare unstrained GaN with GaN@MgSiN$_2$. The reciprocal space
lines $\Gamma -$X, $\Gamma -$Y, and $\Gamma-$Z are aligned with the corresponding Cartesian
directions in real space, shown in Fig.~\ref{fig:compare_vecs}. The strain induces an inversion
of the heavy-hole band and of the split-off hole band along the compression direction
$\Gamma -$Y, leading to a reduced effective mass in the same direction. The energy separation between
the inverted bands along this direction is 50~meV for ZnGeN$_2$, and 320~meV for MgSiN$_2$.
We note that this behavior is only observed for the compression direction, while band inversion
does not occur along the directions of tensile strain ($\Gamma -$X and $\Gamma -$Z).

In Figs.~\ref{fig:bands}(e)-(g) we show maps of the energy of the highest valence band of
GaN, GaN@ZnGeN$_2$, and GaN@MgSiN$_2$ in the $k_x$-$k_y$ plane, for $k_z=0$. Panel (e)
shows the expected six-fold symmetry of the band of unstrained GaN. Panels (f) and (g)
show that strain reduces this symmetry and the map adopts the symmetries of the orthorhombic
structure, with the steepest dispersion and lightest mass along the $k_y$ direction.

To quantify the change in effective masses, we calculate the curvature of the GW bands along
the $k_x$, $k_y$, and $k_z$ directions using finite differences within 5 meV of the valence-band maximum. The results of these calculations
are reported in Table~\ref{tab:masses}, and methodological details are provided in the supplementary material.
Unstrained GaN exhibits an effective mass $m^*=0.48~m_e$ along the $\Gamma -$X and $\Gamma -$Y directions, and $m^*=1.98~m_e$ along $\Gamma -$Z.
These values are in agreement with previous GW calculations\cite{Ponce:2019}. In the case of anisotropically strained GaN, we find lighter effective masses along the direction of compression, namely $m^* = 0.21~m_e$ and $m^* = 0.18~m_e$ for GaN@ZnGeN$_2$ and GaN@MgSiN$_2$.
The electron effective masses can be found in Table S4 in the supplementary material.

\begin{table}
    \centering
    \caption{
Directional hole effective masses of GaN, \ganz, and \ganm. We report the hole effective masses for the
highest valence band in each case. This mass corresponds to the heavy holes for GaN, and to the
split-off holes for \ganz\ and \ganm.
All masses are in units of the free electron mass.}
    \label{tab:masses}
    \begin{ruledtabular}
\begin{tabular}{lcccc}
& Direction & GaN & \ganz\ & \ganm\ \\[3pt]
\hline\\[-8pt]
Holes  & $x$ & 0.48 & 1.57 & 1.62 \\
& $y$ & 0.48 & 0.21 & 0.18 \\
& $z$ & 1.98 & 2.08 & 2.30 \\[1pt]
\end{tabular}
\end{ruledtabular}
\end{table}

We note that we find significant non-parabolicity in the valence band frontier of GaN, as noted in previous theoretical calculations\cite{Rinke:2008}.
In particular, a parabolic dispersion is found only within a 5~meV energy range from the valence band top in unstrained GaN. Below this energy, a band swap occurs and the heavy-hole band becomes the highest valence band.
The valence bands of strained \ganz\ also display non-parabolicity: the bands within 50 meV of the maximum are split-off holes with low effective masses, while the bands below 50 meV are the heavy holes.
A detailed analysis of the variation of the effective mass with energy is reported in Fig.~S2.

We also checked the effect of strain on the phonon dispersion relations of GaN. Fig.~S3 shows the change
of the phonon band structures from unstrained GaN to GaN@ZnGeN$_2$ and GaN@MgSiN$_2$. As expected,
the only significant changes are observed for GaN@MgSiN$_2$, which carries the largest strain.
In this case we find an 11\% reduction in group velocity of transverse acoustic phonons along the
compression direction, and a 2.5\% increase in the longitudinal optical phonon energy.

\begin{figure}
\includegraphics[width=1.0\columnwidth]{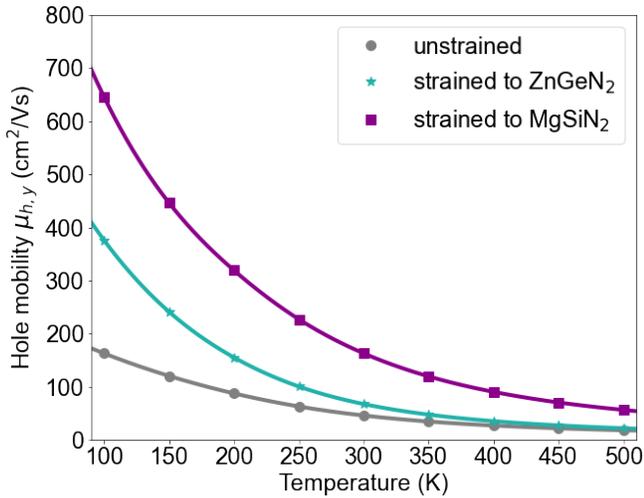}
\caption{\label{fig:mobs}
Hole mobility in the $y$ direction as a function of temperature, calculated with the full \bte, for unstrained
GaN (grey), \ganz\ (teal), and \ganm\ (magenta).
}
\end{figure}

Our \bte\ calculations yield room-temperature hole mobilities along the $x$ and $y$ directions $\mu_x = \mu_y = 46$~\cms, and $z$ direction $\mu_z = 42$~\cms\ for unstrained GaN, in agreement with previous calculations\cite{Ponce:2019}.
Our results slightly exceed the mobility of 31~\cms\ measured on $p$-type, ultra-low defect-density GaN samples\cite{Horita:2017}.
Upon application of strain, the mobility increases to $\mu_y = 67$~\cms\ for GaN@ZnGeN$_2$
(47\% increase) and to $\mu_y = 163$~\cms\ for GaN@MgSiN$_2$ (255\% increase).
Temperature-dependent mobilities for all three cases are shown in Fig.~\ref{fig:mobs}.

Along the $x$ and $z$ directions, where the material experiences tensile strain, the hole mobility is only slightly changed: in the case of \ganz, the mobility decreases by 12\% along the $x$ direction and by 1\% along the $z$ direction; in the case of \ganm, the mobility increases by 4\%  along $x$ and decreases by 7\% along $z$.
The temperature-dependent mobilities along the $x$ and $z$ directions are shown in Figs.~S4 and S5.
We have also analyzed electron mobilities, but due to the largely isotropic nature of the conduction band, the values for \ganz and \ganm\ show only minor deviations from the values in unstrained GaN (see supplementary material).

The mobility is mostly dictated by the band dispersions, which have been discussed above, and the carrier scattering rates. We analyze the most representative electron-phonon scattering rates\cite{Ponce:2019_3} by considering holes located at $3k_{\rm B}T/2=39$~meV below the valence-band maximum, with $T=300$~K. To do so, we perform a spectral decomposition of the angular-averaged scattering rate by phonon energy (see Fig.~S6). For unstrained and strained GaN the trends are consistent: we find that 80\% of the scattering rate originates from acoustic phonon scattering, and in particular piezoacoustic scattering. Scattering by longitudinal-optical phonons account for 15\% of the total scattering rate, and scattering by transverse-optical phonons accounts for the remaining 5\%. The scattering rates as a function of energy from the valence band edge are reported in Supplemental Fig.~S7, and their values at $3k_{\rm B}T/2$ are reported in Table~\ref{tab:mobs}. We see that the scattering rates are much less sensitive to strain than the effective masses, therefore the main source of mobility enhancement is the decrease in the hole mass.


It is instructive to compare our mobility in the compressive $y$ direction from \bte\ calculations with the trends that one obtains from a simple Drude model, with values reported in Table \ref{tab:mobs}.
While the trends in the Drude mobilities are similar to the \bte\ results, the Drude values are off by a factor of two.
We attribute this to the fact that the valence-band structures of \ganz\ and \ganm\ display significant anisotropy and a complex, non-parabolic structure; a first-principles treatment is therefore required to accurately describe hole transport in strained GaN systems.

\begin{table}
    \centering
    \caption{
Effective masses $m^{*}_{y}$, scattering rates $1/\tau$, and mobilities in the compressive strain direction ($y$) for holes in GaN, \ganz, and {\ganm}.
The scattering rates are evaluated at room temperature for holes with energy $3k_{\rm B}T/2$ below the top of the valence band.
The Drude mobility $\mu_{h,y}$=$e\tau/m_{h,y}^{*}$ is only an approximation, and is compared with
the accurate \bte\ mobility resulting from full Boltzmann transport calculations including all electron-phonon scattering processes.
}
    \label{tab:mobs}
    \begin{ruledtabular}
\begin{tabular}{c c c c c }
&  & GaN & \ganz\ & \ganm\ \\[2pt]
  \hline \\[-6pt]
  $m^{*}_{y}$ ($m_e$) & & 0.48 & 0.21 & 0.18 \\
 $1 / \tau$ (THz) & & 54.0 & 59.2 & 43.9 \\[2pt]
  \hline \\[-6pt]
  $\mu_{h,y}$ (\cms) & Drude & 68 & 141 & 223 \\
& \bte\ & 46 & 67 & 163
\end{tabular}
\end{ruledtabular}
\end{table}

\begin{figure}
\includegraphics[width=0.99\columnwidth]{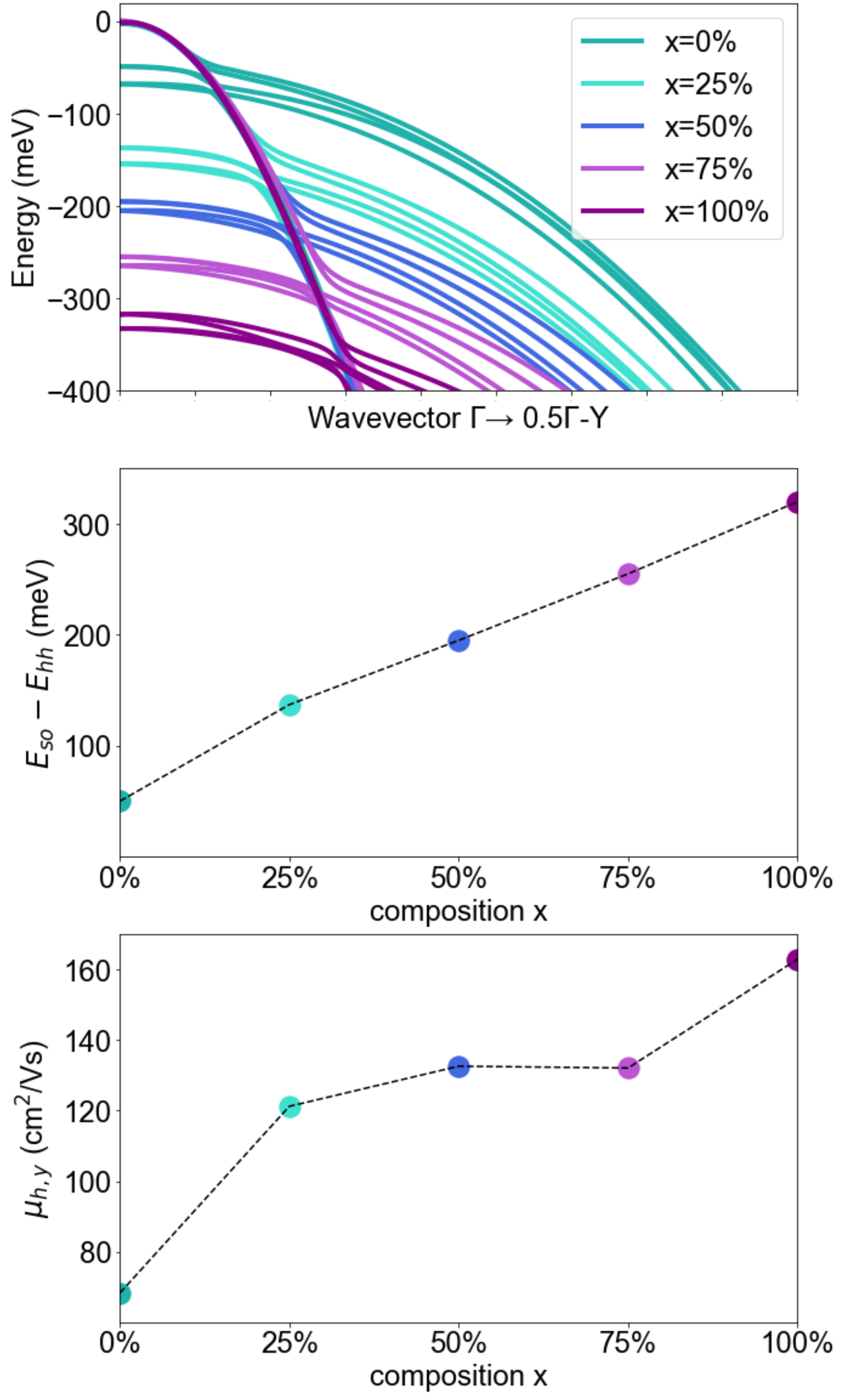}
\caption{\label{fig:alloy}
(a) Valence-band structures of GaN grown on hypothetical solid solutions of
Zn$_{1-x}$Ge$_{1-x}$Mg$_x$Si$_x$N$_2$, which correspond to ZnGeN$_2$ and MgSiN$_2$ in the limiting
cases of $x=0$ and $x=1$, respectively. (b) Energy separation between the split-off hole band and
the heavy hole band at the zone center, as a function of $x$ in Zn$_{1-x}$Ge$_{1-x}$Mg$_x$Si$_x$N$_2$.
(c) Calculated room-temperature hole mobility of strained GaN along the $y$ direction, as a function of alloy composition $x$.
}
\end{figure}

The strain values for \ganz\ are modest and should allow for pseudomorphic growth of GaN layers of sufficient thickness for device structures.  The \ganm\ values are probably too large to allow for growth of device structures; we are including MgSiN$_2$ to establish trends, and in the eventuality that partially relaxed or alloyed layers based on MgSiN$_2$ could be used in structures with a smaller mismatch.
Given that the mobility enhancement in \ganz\ is modest, and that the more appealing \ganm\ may be difficult to realize due to the high strain, we also considered the possibility of modifying the lattice parameters of ZnGeN$_2$ through alloying.
In this context, we assess the potential of hypothetical solid solutions of Zn$_{1-x}$Ge$_{1-x}$Mg$_x$Si$_x$N$_2$.
Figure~\ref{fig:alloy}(a) shows how the band structure of GaN evolves under increasing strain from $x=0$ to $x=1$.
As expected, with increasing compressive strain the heavy hole band is further lowered in energy, thereby increasing the energy separation between the split-off holes and the heavy holes, as shown in Fig.~\ref{fig:alloy}(b).
However, while the energy separation between split-off holes and heavy holes is linear in the strain, we find that the hole mobility varies nonlinearly, and tends to saturate near $x=0.25$.
At $x=0.25$ the hole mobility in the $y$ direction is already as high as 121~\cms, which is more than 160\% higher than in unstrained GaN.

Finally, the effects of including the long-range quadruple moment, omitted in this work, in interpolating the electron-phonon matrix elements in GaN have been previously explored\cite{Jhalani:2020,Park:2020,Brunin:2020,Ponce:2021}. 
It was shown that the mobility of GaN increases from 45~cm$^{2}$/Vs to 55~cm$^{2}$/Vs when including quadrupole corrections (22\% increase). 
We demonstrate that the scattering physics across strain states remains consistent, and the increase in hole mobilities is driven by the change in band dispersions under strain.  
The inclusion of the long-range quadruple would thus scale the mobilities upwards, but we anticipate the trends with strain to remain the same.  
We have additionally tested the influence of including long-range quadruples moments in the interpolation on the calculated hole mobility $\mu_{y}$ in unstrained GaN and found an increase to 53 cm$^2$/Vs from 47 cm$^2$/Vs, in agreement with previous studies.

In summary, using state-of-the-art first-principles calculations and the \textit{ab initio} Boltzmann transport equation, we investigated
the possibility of increasing the room-temperature hole mobility of GaN through anisotropic in-plane strain via pseudomorphic epitaxial growth on ZnGeN$_2$
and MgSiN$_2$. We found that \ganz\ can lead to an increase in the hole mobility along the direction of compression
from 46~\cms\ in unstrained GaN to 76~\cms, and \ganm\ can increase the mobility up to 163~\cms.
Due to the nonlinear dependence of mobility on strain, a hypothetical solid solution of 25\% MgSiN$_2$
added to ZnGeN$_2$ would already increase the hole mobility to 121~\cms.
Beyond strain engineering $p$-channel GaN devices, growth of effectively uniaxial compressed GaN on II-IV nitrides could pose a promising route for engineering monolithic GaN complementary metal–oxide–semiconductor (CMOS) devices. 
We hope that the present study will motivate renewed experimental efforts toward the realization
of high-mobility $p$-type GaN.

This research is supported by the Computational Materials Sciences Program funded by the U.S.
Department of Energy (DOE), Office of Science, Basic Energy Sciences BES), under Award No. DE-SC0020129
(J.L.: calculations and data analysis, manuscript preparation; F.G.: project supervision, manuscript preparation).
S.P. acknowledges support from the F.R.S.-FNRS and the Consortium des \'Equipements de Calcul Intensif (C\'ECI).
Work by N.L.A. was supported by the Army Research Office (W911NF-16-1-0538).
Work by C.G.VdW. was supported by the US DOE, Office of Science, BES, under Award No. DE-SC0010689.
The authors acknowledge the Texas Advanced Computing Center (TACC) at The University of Texas at
Austin for providing HPC resources, including the Frontera and Lonestar5 systems, that have
contributed to the research results reported within this paper. URL: http://www.tacc.utexas.edu.
This research used resources of the National Energy Research
Scientific Computing Center, a DOE Office of Science User Facility supported by the Office of
Science of the U.S. Department of Energy under Contract No. DE-AC02-05CH11231.
Finally, this material is based upon work supported by the National Science Foundation under OAC Grant No. 2103991 (software development, J.L.).

\section{Supplementary Material}

Supporting information, including full calculation details and additional data, can be found in the Supplementary Materials document. 

\section*{Conflicts of Interest}

The authors have no conflicts of interest to disclose

\section*{Data Availability Statement}

The data that support the findings of this study are available from the corresponding author upon reasonable request.

\nocite{*}
\bibliography{references}

\end{document}